\documentclass[useAMS,usenatbib]{mn2e}
\usepackage{amssymb}

\usepackage{pstricks}
\usepackage{color}
\usepackage{graphicx}
\usepackage{slashed}
\usepackage{relsize}
\begin{document}

\title[Microlensing pulsars]
{Microlensing pulsars}

\author[Dai, Xu]{S. Dai$^1$, R. X. Xu$^1$ and A. Esamdin$^2$\\
$^1$School of Physics and State Key Laboratory of Nuclear Physics
and Technology, Peking University, Beijing 100871, China\\
$^2$Urumqi Observatory, National Astronomical Observatories, 40-5
South Beijing Road, Urumqi, 830011, China\\}

\maketitle

\begin{abstract}
We investigate the possibilities that pulsars act as the lens in
gravitational microlensing events towards the galactic bulge or a
spiral arm.
Our estimation is based on expectant survey and observations of
FAST (Five hundred meter Aperture Spherical Telescope) and SKA
(Square Kilometer Array), and two different models of pulsar
distribution are used.
We find that the lensing rate is $\geq 1$ event/decade, being high
enough to search the real events. Therefore, the microlensing
observations focusing on pulsars identified by FAST or SKA in
the future are meaningful.
As an independent determination of pulsar mass, a future detection
of microlensing pulsars should be significant in the history of
studying pulsars, especially in constraining the state of matter
(either hadronic or quark matter) at supra-nuclear densities.
The observations of such events by using advanced optical facilities
(e.g., the James Webb Space Telescope and the Thirty Meter
Telescope) in future are highly suggested.
\end{abstract}


\section{Introduction}

Pulsars could be normal neutron stars or quark
stars~\citep{Lattimer2004}, and it should be very important to
affirm or negate the existence of either neutron or quark stars in
order to guide physicists in studying the nature of fundamental
strong interaction.
With regard to the possible ways of identifying quark
stars~\citep[e.g.,][for a review]{Xu08}, it could be very
straightforward and clear if we would find low-mass quark stars,
since neutron stars are essentially gravitation-bound while low-mass
quark stars are mainly confined by strong interaction.
If we can detect a pulsar-like star with mass $\lesssim 0.1M_\odot$,
then it is surely a quark star.

How to measure the mass of stars?
Up to now only the masses of compact stars in binary systems
have been determined, by either Keplerian or
post-Keplerian parameters.
The lowest mass detected of an eclipsing X-ray pulsar, SMC X-1,
is $1.06_{-0.10}^{+0.11}~M_\odot$, which is near the minimum
mass expected for a neutron star produced in a supernova~\citep{LMC
X-1}.
However, most of pulsars are isolated, and it is still a big
challenge to measure their masses. It was suggested to
observationally determine an isolated neutron star's mass from the
red-shift (as a function of the ratio of mass to radius, $M/R$) and
pressure broadening (as a function of $M/R^2$) of an absorption
spectrum, yet no atomic line has been detected with certainty in the
thermal X-ray spectra~\citep[e.g.,][]{Xu02}.

Interestingly, gravitational microlensing~\citep[e.g.,][]{mao},
which makes use of the temporal brightening of a background star due
to intervening object, would provide us a powerful method to measure
the masses of compact objects. The amplification $A$ of a background
star by the passage of a pulsar has a relation of
$$
A=\frac{u^{2}+2}{u\sqrt{u^{2}+4}},
$$
where $u=r'/R_{\rm E}$ is a parameter, with the undeflected distance, 
$r'$, and the Einstein radius, $R_{\rm E}$ ~\citep{Paczy}, which is
almost a function of pulsar mass $M$ and $r$ if the background
star's distance $D\gg r$ [see Eq.(\ref{RE})]. One can then directly
obtain the mass-distribution of the lens objects via the light
curve~\citep{Glicenstein}.
With the method of microlensing, we can not only determine the mass
distribution of massive compact halo
objects~\citep[MACHO,][]{Alcock} and massive objects in the Galactic
center region~\citep{Gil}, but may also measure the masses of
isolated pulsars.

The possibilities of microlensing neutron star have been discussed
previously and several estimation of lensing rates had been
presented in the literatures~\citep{Horvath,Schwarz}. In this paper
we reestimate the lensing rate by calculating the solid angle
Einstein ring swept due to pulsar's proper motion, using two different
pulsar distribution models.
In addition, our estimation also takes expectant observations of
FAST~\citep{Nan} and SKA~\citep{Johnston2007} into account, since
the two new telescopes, FAST and SKA, will greatly enhance our
ability of searching and observing pulsars, and the number of pulsars
with measured proper motion in the Galaxy is crucial to estimate the
lensing rate. Lorimer et al. (2006) used the results from recent
surveys with the Parkes Multi-beam system to derive a potentially
detectable population of $30000$ normal pulsars. By using this number
and assuming $30000$ potentially detectable millisecond pulsars in the
Ga1axy~\citep{Lyne}, Smits et al. (2009a) estimated that the all-sky
survey with only the $1$-km core of the SKA located in the southern
hemispheric would detect about $14000$ normal pulsars and about $6000$
millisecond pulsars. Under the same assumption, Smits et al. (2009b)
noted that more than $7000$ previously unknown pulsars could be discovered
by FAST in the Galactic plane ($|b|<10^{\rm o}$) with an observation
time of $1800$ s per-beam. Further more, by regular timing these
pulsars using FAST and SKA, the accurate positions and proper motion
of them can also be determined. We then propose to search possible
microlensing events due to pulsars identified by FAST and SKA and
to carry out microlensing observations coupled with radio observations
by means of future optical projects (e.g., the James Webb Space
Telescope\footnote{http://www.jwst.nasa.gov/} or the Thirty Meter
Telescope Project\footnote{http://www.tmt.org/}).

The lensing rate of our result indicates a high possibility of
observing microlensing event due to pulsars in the future with the
help of FAST and SKA.
We expect an independent measurement of pulsar mass through these
microlensing observations, especially to discovery low-mass pulsars,
in order to understand the real nature of pulsars.

\section{The lensing rate}

We consider a pulsar at distance $r$ with velocity $\upsilon$, microlensing
a background star at distance $D$. The Einstein ring of this pulsar
sweeps a solid angle $S_{\rm N}$ on the celestial sphere during a period
time of $t$.
The Einstein radius is
\begin{equation}R_{\rm E}=\sqrt{\frac{2R_{\rm S}}{D}r(D-r)},
\label{RE}%
\end{equation}
where $R_{\rm S}=2GM/c^2$ denotes the Schwarzschild radius.
This solid angle $S_{\rm N}$ depends on the mass $M$, the distance
$r$ to the pulsar and also the distance $D$ to the star,
\begin{equation}S_{\rm N}(M,\upsilon,r,D)=\frac{R_{\rm E}}{r}\frac{\upsilon t}{r}.\end{equation}

We expect that a microlensing event occurs when a star in the
background falls into this solid angle. We assume that our telescopes
can identify a total of $N$ pulsars, and $S$ square degrees of the
Milky Way is visible to it, then the lensing rate per unit time can
be estimated as
\begin{equation}p=\frac{\sum\limits_{\rm N}S_{\rm N}}{S}N_{\rm star},\end{equation}
where $N_{\rm star}$ denotes the total number of stars visible.

If we take pulsar distribution into account, the sum turns
into multiple integral on the space of position ($r$ and $D$)
and velocity ($\bf \vec \upsilon$). Then the lensing rate $p$
can be rewritten as
\begin{equation}p=\frac{\int_{\rm position}\int_{\rm velocity} S_{\rm N}({\bf \vec \upsilon}, \vec{r}, \vec{D}){\rm d}{\bf \vec \upsilon}{\rm d}\vec{r}{\rm d}\vec{D}}{S}N_{\rm star},\end{equation}
where $D\geq r$. We assume that the number density of stars in the
galactic bulge or a spiral arm is constant in following estimation.
Because the velocity distribution of pulsars is not a function of
$r$ and $D$, one can integrate the velocity distribution independently
of position distribution. We can thus just apply a mean
velocity in estimating the lensing rate. Hansen and Phinney (1997)
suggested a Maxwellian distribution for the kick velocities of
pulsars,
\begin{equation}f(\upsilon_{\rm kick})=\sqrt{\frac{2}{\pi}}\frac{\upsilon_{\rm kick}^{2}}{\sigma^{3}}\exp(-\frac{\upsilon_{\rm kick}^{2}}{2\sigma^{2}}),\end{equation}
where $\sigma=190$ $\rm{km/s}$ and mean kick velocity $\upsilon_{\rm kick}=300$
$\rm{km/s}$ are chosen. Considering that only the part of the
velocity vector that lies in the lens plane is effective, one should
use the mean velocity projection on the lens plane~\citep{Schwarz},
\begin{equation}\upsilon=\frac{1}{\pi}\int_{-\pi/2}^{\pi/2}\upsilon_{\rm kick}\cos\alpha {\rm d}\alpha,\end{equation}
where the projected velocity, $\upsilon$, is order of $200$ $\rm{km/s}$.

We use two different models of pulsar number density from Hartman et
al. (1997), which resemble the models from Narayan (1987) and
Johnston (1994), see also Schwarz and Seidel (2002), in our
following simulations.
The first model of pulsar distribution is
\begin{equation}n_{\rm P1}(R)=\frac{1}{2\pi R_{\rm W}^{2}}\exp (-\frac{R}{R_{\rm W}}),
\label{np1}%
\end{equation}
where $R$ is the radial distance of the pulsar to the galactic
center in the galactic plane, and $R_{\rm W}=5$ kpc.
The second model is
\begin{equation}n_{\rm P2}(R)=\frac{c_{\rm P2}}{2\pi R_{\rm W}^{2}}\exp (-\frac{(R-R_{\rm max})^{2}}{2R_{\rm W}^{2}}),
\label{np2}%
\end{equation}
where $R_{\rm W}=1.8$ kpc and $R_{\rm max}=3.5$ kpc.
The normalization constant is $c_{\rm P2}=0.204$ for the given choice of $R_{\rm max}$.
For the $z$-dependence we apply
\begin{equation}n_{z}(z)=\frac{1}{\sqrt{2\pi}\sigma}\exp (-\frac{1}{2}\frac{z^{2}}{\sigma^{2}}),\end{equation}
with $\sigma=0.45$ kpc~\citep{Lyne, Schwarz}.

We transform the pulsar distribution into the spherical coordinates
$(r, \theta, \phi)$ with the sun at the origin via
\begin{equation}z=r\sin\theta,\end{equation}
\begin{equation}R^{2}=r^{2}\cos^{2}\theta+R_{\rm SC}^{2}-2rR_{\rm SC}\cos\theta \cos\phi,\end{equation}
where $R$ is the radial distance of the pulsar
to the galactic center in the galactic plane
and $z$ is the height, and $R_{\rm SC}=8.5$ kpc.
For pulsar distribution $n_{\rm P1}$, we have

\[
p=\frac{NN_{\rm star}}{S}\int\int\int n_{\rm P1}(R)n_{z}(z)r^{2}\sin\theta {\rm d}r{\rm d}\theta {\rm d}\phi\
\]

\[
\times \frac{1}{r^{2}}\frac{\upsilon t}{4\pi r}\int_{0}^{r}\sqrt{\frac{2R_{\rm S}}{D}r(D-r)}{\rm d}D
\]

\[
=\frac{NN_{\rm star}\upsilon t}{2\pi S R_{\rm W}^{2}\sqrt{2\pi}\sigma}\int\int\int
\]

\[
\exp(-\frac{\sqrt{r^{2}\cos^{2}\theta+R_{\rm SC}^{2}-2rR_{\rm SC}\cos\theta \cos\phi}}{R_{\rm W}})
\]

\begin{equation}\exp(-\frac{1}{2}\frac{r^{2}\sin^{2}\theta}{\sigma^{2}})\sin\theta {\rm d}r{\rm d}\theta {\rm d}\phi\frac{\sqrt{2R_{\rm S}r}}{4\pi}\int_{0}^{1}\sqrt{\frac{y-1}{y}}{\rm d}y,\end{equation}
where $y=D/r$.
If we define $x=r/R_{\rm SC}$, then
the expression can be reduced to
\[
p=\frac{NN_{\rm star}\upsilon t R_{\rm SC}\sqrt{2R_{\rm S}R_{\rm SC}}}{8\pi^{2} S R_{\rm W}^{2}\sqrt{2\pi}\sigma}
\]

\[
\int\int\int\exp(-\frac{\sqrt{x^{2}\cos^{2}\theta+1-2x\cos\theta \cos\phi}}{R_{\rm W}/R_{\rm SC}})
\]

\begin{equation}\exp(-\frac{1}{2}\frac{x^{2}\sin^{2}\theta R_{\rm SC}^{2}}{\sigma^{2}})\sin\theta {\rm d}x{\rm d}\theta {\rm d}\phi\int_{0}^{1}\sqrt{\frac{y-1}{y}}{\rm d}y.
\label{p}%
\end{equation}

We consider that FAST is expected to detect about $7770$ pulsars in
about $70\times10$ square degrees of the Milky Way of observing time
in less than a year~\citep{Nan}, while SKA is expected to detect about
$15000$ pulsars in $290\times10$ square degrees of the Milky Way.
If one chooses $M=1M_\odot$, $r\in[0,5 \rm{kpc}]$, $t = 1$ yr, $N=15000$,
$N_{\rm star}=10^{11}$, $\upsilon=200$ $\rm{km/s}$, and $S=2000$ square
degrees, then we have $p\approx12$ event/year from Eq.(\ref{p}) .

Certainly the total number of stars visible is less than $10^{11}$
if the effect of extinction is included. Based on the photometric maps
of the galactic bulge released by the OGLE (Optical Gravitational Lensing
Experiment) project, which contain photometry of about $30$ million stars
from $49$ fields covering $11$ square degrees in different regions of the
galactic bulge~\citep{Udalski}, we estimate that in the region of $S=2000$
square degrees, $10^{9}$ stars would be visible, then the lensing rate
should be $p\geq 1$ event/decade according to Eq.(\ref{p}).
The lensing rate density as function of $\theta$ and $\phi$ is shown
in Fig.$1$.

%
\begin{figure}
  \includegraphics[width=3 in]{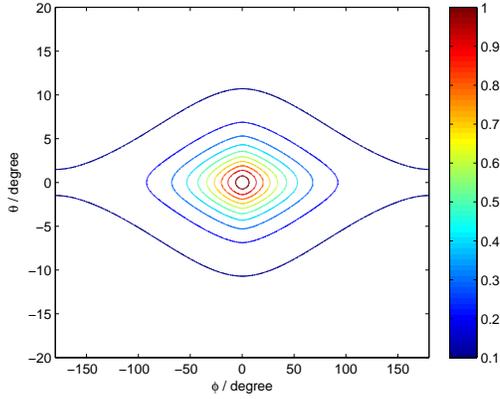}
\caption{The lensing rate density as function of $\theta$ and $\phi$ in the
model of pulsar distribution described by Eq.(\ref{np1}), where
$\theta$ is the azimuth angle and $\phi$ is the polar angle. Color bar stands for
lensing rate in units of number of events/decade/${\rm degree}^{2}$.}
\end{figure}

For pulsar distribution $n_{\rm P2}$ in the model of pulsar
distribution described by Eq.(\ref{np2}), we have accordingly
\[
p=\frac{NN_{\rm star}}{S}\int\int\int n_{\rm P2}(R)n_{z}(z)r^{2}\sin\theta {\rm d}r{\rm d}\theta {\rm d}\phi
\]

\[
\times \frac{1}{r^{2}}\frac{\upsilon t}{4\pi r}\int_{0}^{r}\sqrt{\frac{2R_{\rm S}}{D}r(D-r)}{\rm d}D
\]

\[
=\frac{NN_{\rm star}c_{\rm P2}\upsilon t}{2\pi S R_{\rm W}^{2}\sqrt{2\pi}\sigma}\int\int\int
\]

\[
\exp(-\frac{(\sqrt{r^{2}\cos^{2}\theta+R_{\rm SC}^{2}-2rR_{\rm SC}\cos\theta \cos\phi}-R_{\rm max})^{2}}{2R_{\rm W}^{2}})
\]

\begin{equation}\exp(-\frac{1}{2}\frac{r^{2}\sin^{2}\theta}{\sigma^{2}})\sin\theta {\rm d}r{\rm d}\theta {\rm d}\phi\frac{\sqrt{2R_{\rm S}r}}{4\pi}\int_{0}^{1}\sqrt{\frac{y-1}{y}}{\rm d}y.\end{equation}
From the same process as for $n_{\rm P1}$ of Eq.(\ref{np1}), then we
have $p\geq2\rm$ event/decade if one chooses $M=1M_\odot$,
$r\in[0,5\rm{kpc}]$, $t = 1$ yr, $N=15000$, $N_{\rm star}=10^{11}$,
$\upsilon=200$ $\rm{km/s}$, $S=2000$ square degrees, and takes the effect
of extinction into account.
The corresponding lensing rate density as function of $\theta$ and $\phi$ is
shown in Fig.$2$.
%
\begin{figure}
  \includegraphics[width=3 in]{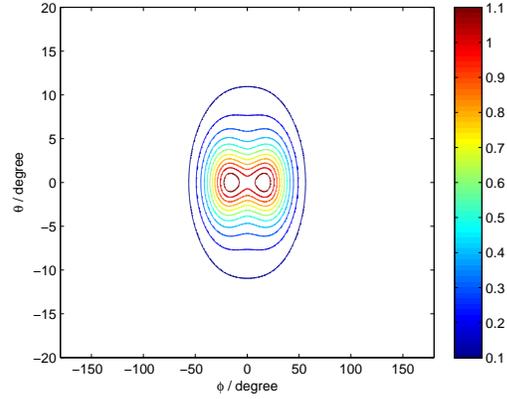}
\caption{Same as in Fig. 1, but in the model of pulsar distribution
described by Eq.(\ref{np2}).}
\end{figure}

The lensing rates estimated above are for the microlensing events in the
galactic center. Considering that the observation of FAST is limited to
the spatial arms and that pulsars mainly distribute on the disk of the
Milky Way, we then transform the pulsar distribution into cylindrical
coordinates with sun at the origin via
\begin{equation}R^{2}=r^{2}+R_{\rm SC}^{2}-2rR_{\rm SC}cos\phi.\end{equation}
Using the cylindrical coordinates, we have
\[
p=\frac{NN_{\rm star}}{S}\int\int n(r,\phi)n_{z}(r)rdrd\phi\times \frac{1}{r^{2}}\frac{\upsilon t}{r}
\]

\begin{equation}\int_{0}^{r}\sqrt{\frac{2R_{\rm S}}{D}r(D-r)}dD.\end{equation}

For pulsar distributions of both $n_{\rm P1}$ and $n_{\rm P2}$, we
calculate following the way we did previously and choose
$M=1M_\odot$, $r\in[0,5\rm{kpc}]$, $t = 1$yr, $N=10000$, $\upsilon=200$ $\rm{km/s}$,
and $S=2000$ square degrees. Considering the effect of extinction,
we choose $N_{\rm star}=10^{9}$. Finally we can also obtain a lensing
probability of $p\geq1$ event/decade for FAST.

\section{A proposal of discovering microlensing pulsars}
We have estimated the probability of observing microlensing pulsars
by means of FAST and SKA. Besides the encouraging lensing rate
according to our results, a feasible searching strategy is also
essential for discovering a microlensing pulsar. We therefore
propose to carry out microlensing pulsar observation coupled with
FAST and SKA observations in the future, which are hopeful to
determine the mass of isolated pulsars.

Microlensing pulsar observation should be based on the expectant
data of FAST and SKA. We propose to systematically list the information of
new pulsars discovered by FAST and SKA in the future, including the position,
proper motion and distance, and high proper motion pulsars should be especially
focused on. The position of these pulsars in the next few decades could also
be predicted. Then we could compare the predicted position of these pulsars
with the position of stars in the galactic bulge and spiral arms, and background
star candidates of microlensing events whose position would be quite close to
the predicted position of pulsars could be picked up. With these candidates,
we should formulate a plan to monitor certain background stars for the predicted
microlensing events in a certain period of time in the future.

Microlensing pulsar observation could be carried out on ongoing
lensing projects, and future advanced facilities are also expected to benefit
the observation. The magnification of a background star could be about $0.3$,
on the condition that a pulsar just arriving at the edge of the Einstein
ring~\citep{mao}, and the time scale of the microlensing event should be about 
$15$ days. The observation can then be done on current telescopes in
the infrared band where the interstellar extinction is much weaker~\citep{Horvath,Udalski}.
Besides ongoing gravitational lensing projects, we could also look forward
to future projects, especially the Thirty Meter Telescope (TMT) Project.
The TMT project, which is scheduled for the next decade, will greatly promote
our study of gravitational lensing on cosmology, galaxy formation and the
distribution of lensing mass. TMT's great capabilities of resolving smaller,
fainter source populations will allow a much higher sky density of background
sources to be used~\citep{Carlberg}, which will directly enhance the possibility
of discovering microlensing pulsar events. Therefore, TMT together with FAST and
SKA will provide us a good opportunity to observe microlensing pulsar events and
determine the mass of isolated pulsars.

We note that the proposed microlensing pulsar observation is costless,
that is to say, it does not need to monitor a target all the time.
Our estimation has shown the possibility of microlensing a pulsar, and
the key of the observation is predicting the microlensing events
based on the large and sensitive database of FAST and SKA, and then
monitoring microlensing candidates of background stars in a predicted short
period of time. Even though the prediction can not be precise and the lensing
rate is relatively small, the costless observation is still meaningful
and should be carried out.

\section{Discussion and conclusions}

We have estimated the lensing rates of pulsars towards the galactic
bulge and spiral arms. Our estimation shows that, for FAST and SKA,
the lensing rates ($\propto N$) are $\geq1$ event/decade at least,
which are much higher than the estimations made in previous
literatures.
The number of pulsars with measured proper motion, $N$, is crucial
to estimate the rate of lensing events. In table $1$, we summarize
the expectant number of pulsars to be detected by FAST and SKA,
based on previous survey simulation results.
In addition, the lensing rate should increase significantly if the
population of rotating radio transients~\citep[RRATs,][]{Keane} is
included, since the total number of RRATs is at least several times
of galactic active radio pulsars~\citep{McLaughlin}.
RRATs, which can be precisely located by timing and dispersion
measurement, should then be one of the key targets of FAST and SKA.
%
\begin{table*}
\begin{center}
\caption{Summary of previous survey simulation results of FAST and
SKA. Numbers in the parentheses represent known pulsars. $l$, $b$
and Dec are the longitude, latitude and declination in the galactic
coordinate system, respectively.}
\begin{tabular}{|l|c|c|c|c|c|}
\hline
      &  Detectable pulsars &         \multicolumn{2}{c|}{FAST$^c$}                 &                 \multicolumn{2}{l|}{~~~~~~~~~~~~~~~~~SKA$^{d}$} \\
\hline
      &  All Sky            &  $20^{\circ}<l<90^{\circ}$  &  $20^{\circ}<l<90^{\circ}$  &  $0^{\circ}<l<85^{\circ}$ \& $155^{\circ}<l<360^{\circ}$ & Dec $< 50^{\circ}$   \\
      &                     &   $|b| \leq 10^{\circ}$     &  $|b| \leq 10^{\circ}$      &  $|b| \leq 5^{\circ}$      &                      \\
\hline
Normal pulsar      & $\sim 30000^a$ & $\sim 5700 (352)$ & $\sim 7000 (418)$ &  $\sim 11000$ &  $\sim 14000$   \\
\hline
Millisecond pulsar & $\sim 30000^b$ & $\sim 550 (14)$   & $\sim 770 (20)$   &  $\sim 4000$  &  $\sim 6000$    \\
\hline
\end{tabular}
\end{center}
Note. -- The results are from ``$a$'':~\cite{Lorimer},
``$b$'':~\cite{Lyne}, ``$c$'':~\cite{Smitsb},
``$d$'':~\cite{Smitsa}.
\end{table*}

We also note that the lensing rates we showed are based on photometric
microlensing events. If we consider astrometric microlensing, which
makes use of the shift of the centroid of the combined images of the
light source, the cross section would increase by a factor of $\sim
10^2$~\citep{Horvath, Schwarz}.
As our lensing rate is proportional to cross section, the
astrometric microlensing probability would be $\sim 10^2$ times
higher than that presented previously in \S2.
It could then be realistic to search astrometric microlensing
pulsars and to detect the events by facilities with very high position
precision in the future.

The lensing rates indicate that it is hopeful to measure the mass of
an isolated pulsar in the future with the method of microlensing. We
propose to do catalogue comparison and microlensing prediction for
pulsars identified by SKA and FAST in the future, and to carry out
microlensing observation coupled with radio observation in order to
detect microlensing pulsar events and to measure the masses of
isolated pulsars with advanced optical facilities.

\section*{Acknowledgments}
\thanks{%
We would like to acknowledge useful discussions at our pulsar group
of PKU, and to thank an anonymous referee for the constructive comments.
This work is supported by the National Natural Science Foundation of 
China (10973002, 10935001), the National Basic Research Program of 
China (2009CB824800), the National Fund for Fostering Talents of Basic 
Science (J0630311) and the Programme of the Light in China¡¯s Western 
Region (LHXZ200602).}

\end{document}